# Comments on "Amplifying magneto-optical photonic crystal, Appl. Phys. Lett. 97, 061116 (2010)"

Amir Djalalian

*Department of Condensed Matter Physics, Royal Institute of Technology, SE-164 40 Kista, Sweden*

## 1. First Comment

In [1], changes to absorption coefficient and refractive index of Erbium doped garnets are attributed to $Er^{3+}$-ion concentration ignoring the effect of host lattice on $Er^{3+}$-ions Stark manifolds. To verify effective absorption coefficient $\alpha(\omega)$ of erbium doped garnets based on above assumptions we proceed as follow. Maximum gain at $\lambda_{res}$ can be expressed as $g = \sigma_e N_2 - \sigma_a N_1 = \sigma_{21} N_2 - \sigma_{12} N_1$ [2] where $N_2+N_1=\rho_0$ is the Erbium ion density, $N_1$ and $N_2$ are the population densities of $Er^{3+}$-ions in $^4I_{15/2}$ and $^4I_{13/2}$ energy states all in $cm^{-3}$. Following the McCumber relation, we can relate the Erbium's absorption to its emission cross section by $\sigma_{21} = \sigma_{12} \exp\left(\hbar \frac{\omega_{21} - \omega}{kT}\right) \rightarrow g = -\sigma_{12}\rho_0 \left\{1 - n_2\left(1 + \exp(\hbar \frac{\omega_{21} - \omega}{kT})\right)\right\}$. Here $n_2=N_2/\rho_0$ is the relative population densities of $Er^{3+}$-ions in $^4I_{13/2}$ energy state and $\omega_{21}$ is a frequency of the resonance transition $2 \rightarrow 1$ at $\lambda_{res}$. Clearly, $\sigma_{12}\rho_0$, hereafter denoted as $\alpha_{Er}(\omega_{21})$, is the maximum signal absorption in a host having Erbium dopant concentration $\rho_0$ when signal overlap factor $\Gamma_s=1$, see equation 1.120 in [2], Thus dispersive relation for gain follows $g_{Er}(\omega) = -\alpha_{Er}(\omega)\left\{1 - n_2\left(1 + \exp(\hbar \frac{\omega_{21} - \omega}{kT})\right)\right\}$. We can then subtract the gain from the absorption of undoped garnet $\alpha_0(\omega)$. Expanding $\alpha=4\pi k/\lambda$ and $\omega=2\pi c/\lambda$ to obtain the effective extinction coefficient $k(\lambda)$ for erbium doped garnets, we get:

$$k(\lambda) = k_0(\lambda) + k_{Er}(\lambda)\left\{1 - n_2\left[1 + \exp\left(\frac{2\pi c\hbar}{kT}\left(\frac{1}{\lambda_{21}} - \frac{1}{\lambda}\right)\right)\right]\right\} - k_{Gd}(\lambda) \quad (1)$$

Similarly, for the real part of the refractive index for Erbium doped garnet we have:

$$n(\lambda) = n_0(\lambda) + n_{Er}(\lambda) - n_{Gd}(\lambda) \quad (2)$$

Here, $n_{Er}$, $k_{Er}$, $n_{Gd}$ and $k_{Gd}$ are the amount of refractive indices and extinction coefficients added/subtracted to/from $n_0$ due to $Er^{3+}$ replacing $Gd^{3+}$ ions. Values for $n_{Er}$, $k_{Er}$ $n_{Gd}$ and $k_{Gd}$ may be calculated from the experimental data available for $Er_2O_3$ and $Gd_2O_3$ in the following way. $Er_2O_3$ has a density=8.64 g/cm$^3$ and a molar mass=382.51 g/mol. Therefore density in terms of number of $Er_2O_3$ per unit volume = density × Avogadro's number /molar mass = $1.359 \times 10^{22}$ cm$^{-3}$ therefore $Er^{3+}$-ion concentration in $Er_2O_3$ is $2 \times 1.359 \times 10^{22} = 2.718 \times 10^{22}$ cm$^{-3}$. In garnets, there are 8 formula units per unit cell. Number of formula units per unit volume for $Gd_3Ga_5O_{12}$ with lattice constant a=12.383 Å [3] is $8/(12.383 \times 10^{-8})^3 = 4.213 \times 10^{21}$ cm$^{-3}$. Assuming that the variation in lattice constant due to Erbium dopant replacing Gadolinium atoms is negligible, $Er^{3+}$-ion concentration in $Gd_2Er_1Ga_5O_{12}$ is $1 \times 4.213 \times 10^{21}$ cm$^{-3}$. The ratio of $Er^{3+}$-ion concentrations in $Gd_2Er_1Ga_5O_{12}$ and $Er_2O_3$ is $\rho'=4.213 \times 10^{21}/2.718 \times 10^{22}=0.156$. $n_{Er}$ and $k_{Er}$ can now be expressed as $n_{Er} = \rho' n_{Er_2O_3}$ and $k_{Er} = \rho' k_{Er_2O_3}$ where complex refractive index for $Er_2O_3$ was determined experimentally in [4]. Replacement of $Gd^{3+}$-ions must be remedied by the same analogy. For the $Gd^{3+}$ replaced by $Er^{3+}$, $\rho''=0.18$ may be calculated based on $Gd_2O_3$ having a density= 7.07 g/cm$^3$ and a molar mass= 362.50 g/mol. For $Gd_{3-x}Er_xGa_5O_{12}$ general expressions for $\rho'=0.156x$ and $\rho''=0.18x$, where x represents Erbium concentration in garnet formula unit. Clearly $(1-\rho') n_0 \neq n_0 - (\rho'' \times n_{Gd2O3})$. Since for undoped $Gd_3Ga_5O_{12}$, $k_0 \approx 0$, equations (1) and (2) may be written as:

$$N_{Er:GGG}(\lambda) = n_{Er:GGG}(\lambda) - ik_{Er:GGG}(\lambda) = \left[ n_0(\lambda) + \rho' n_{Er_2O_3}(\lambda) - \rho'' n_{Gd_2O_3}(\lambda) \right] \\ -i\left[ \rho' k_{Er_2O_3}(\lambda) \left\{ 1 - n_2 \left[ 1 + \exp\left( \frac{2\pi c\hbar}{kT}\left( \frac{1}{\lambda_{res}} - \frac{1}{\lambda} \right) \right) \right] \right\} - \rho'' k_{Gd_2O_3}(\lambda) \right] \quad (3)$$

2. **Second Comment**

Although inclusion of McCumber relation in conjuction of $Er_2O_3$ breaks the dependency on cross sections, one must consider the range of wavelength in which this equation is deemed to be valid. A typical emission and absorption cross section spectra for Erbium has a FWHM of 40-60nm [5] around the resonance wavelength $\lambda_{res}$=1.53μm caused by separation of sublevels in $^4I_{15/2}$ and $^4I_{13/2}$ Stark manifolds. Furthermore both emission and absorption cross sections approach 0 for 1450>λ>1630nm. Therefore 1450<λ<1630nm is the widest possible wavelength range in which equation (2) reported in [1] may be valid. Transmission spectrum depicted in Figure 1 in [1] was only reproducible when the wavelength dependant exponent of McCumber relation was replaced by 1 for all wavelengths, see figure 1b bellow. Inclusion of McCumber relation in its entirety as it was reported in [1] produced a different result at $\lambda > \lambda_{res}$, see figure 1a.

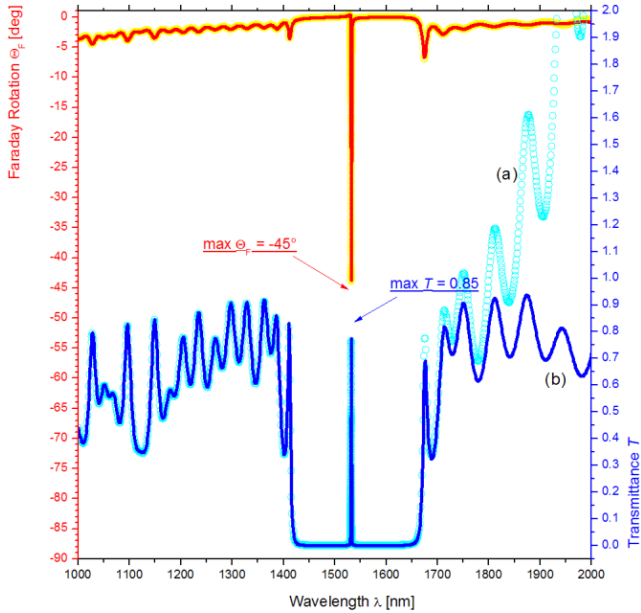

Figure 1: Transmission spectra of multilayer structure reporter in [1] with a) McCumber inclusion. b) When exponent in McCumber relation was replaced with 1.

In summary, Erbium's emission and absorption cross sections vary depending on the host. These variations are due to the Stark sublevels' splitting caused by the lattice, in this case Er:GGG, thus every host has unique finger print on emission and absorption cross sections. Furthermore, comparing the Erbium concentration in $Er_2O_3$ to Er:GGG, one can conclude that even at high dopant concentration, Erbium atoms in GGG are located further apart from each other, resulting in a less likelihood of intermixing of energy levels with those of the neighbouring dopant atoms. Therefore, energy levels of $Er^{3+}$-ions in GGG are perceived to be more discrete compared to those in $Er_2O_3$, which also has a direct impact on the cross sections. Although, in absence of experimental data for Erbium's cross sections in GGG, inclusion of McCumber relation in conjunction with $Er_2O_3$ experimental data may provide us with an approximation, an accurate model is only possible when Erbium cross sections in GGG are measured and identified experimentally.